\documentclass[pre,superscriptaddress,twocolumn,final,showpacs,showkeys,nobibnotes,titlepage,twoside,10pt]{revtex4-1}
\usepackage{amsmath}
\usepackage{graphicx}
\usepackage{amsfonts}
\usepackage{amssymb}
\usepackage{subfigure}
\usepackage{float}
\usepackage{color}

\begin{document}

\author{D. V. Denisov}
\affiliation{Van der Waals-Zeeman Institute, University of Amsterdam, The
Netherlands}
\author{M. T. Dang}
\affiliation{Van der Waals-Zeeman Institute, University of Amsterdam, The
Netherlands}
\author{B. Struth}
\affiliation{Deutsches Elektronen-Synchrotron, Hamburg, Germany}
\author{A. Zaccone}
\affiliation{Physics Department and Institute for Advanced Study, Technische Universit\"{a}t M\"{u}nchen, 85748 Garching, Germany}
\author{G. H. Wegdam}
\affiliation{Van der Waals-Zeeman Institute, University of Amsterdam, The
Netherlands}
\author{P. Schall}
\affiliation{Van der Waals-Zeeman Institute, University of Amsterdam, The
Netherlands}

\title{Non-equilibrium first order transition marks the mechanical failure of glasses}

\begin{abstract}
Glasses acquire their solid-like properties by cooling from the supercooled liquid via a continuous transition known as the glass transition. Recent research on soft glasses indicates that besides temperature, another route to liquify glasses is by application of stress that forces relaxation and flow. Here we provide experimental evidence that the stress-induced onset of flow of glasses occurs via a sharp first order-like transition. Using simultaneous x-ray scattering during the oscillatory rheology of a colloidal glass, we identify a sharp symmetry change from anisotropic solid to isotropic liquid structure at the transition from the linear to the nonlinear regime. Concomitantly, intensity fluctuations sharply acquire liquid distributions.
These observations identify the yielding of glasses to increasing stress as sharp affine-to-nonaffine transition, providing a new conceptual paradigm of the yielding of this technologically important class of materials, and offering new perspectives on the glass transition.
\end{abstract}

\maketitle

Glasses acquire their solid-like properties upon cooling from supercooled liquids at the glass transition~\cite{Ediger1996}: the microscopic relaxation time increases rapidly, but continuously upon approaching the glass transition from the liquid. Despite extensive searches for underlying phase transitions, these have never been observed experimentally. Microscopic correlation length remain limited to a few particle diameters.
Recent research on soft glasses such as colloids, emulsions and foams, have provided another route to liquify a glass by the application of stress. The applied stress forces relaxation and eventually induces flow when exceeding the yield stress. This leads to loss of the elastic properties of glasses and irreversible flow. This failure to mechanical stress is ubiquitous to all amorphous solids and severely limits their applications; yet, its understanding remains challenging. Non-affine particle displacements i.e. deviations from the affine elastic deformation field are believed to be central to the onset of flow, but their understanding remains incomplete.

Recent experiments on soft glasses reveal long-range correlations in the microscopic flow of glasses~\cite{chikkadi_schall11,lemaitre2009}. Such long-range correlations reflect a high susceptibility of the microscopic dynamics to the applied shear. This is in contrast to the glass transition, where dynamic correlations are limited to a few atomic diameters~\cite{BiroliBouchaud}. These long-range correlations should affect the way glasses respond to external stress
however, understanding their effect on the microscopic yielding of glasses i.e. the onset of irreversible deformation, remains a crucial challenge.

Colloidal glasses provide benchmark system to study the structure and dynamics of glasses at the single-particle level. The constituent particles exhibit dynamic arrest due to crowding at volume fractions above $\phi_g \sim 0.58$, the colloidal glass transition~\cite{Pusey1987,Pusey1986,vanMegen,vanMegen1998}. Microscopically, the particles are trapped within cages formed by their nearest neighbors allowing only for very slow structural rearrangements and leading to glass-like properties such as slow relaxation and aging~\cite{Bouchaud1992}. Their yielding to stress has been widely investigated by oscillatory rheology, in which the sample is probed with a time-dependent, oscillatory strain~\cite{YieldingCollGlass}. Here, the onset of flow is associated with the transition from the linear to the nonlinear response regime, also referred to as "yielding" to the oscillatory strain~\cite{YieldingCollGlass}. However, despite intense macroscopic studies, the microscopic mechanism of this transition has remained elusive.
Recent experiments on attractive gel systems \cite{Moller2008,Bartolo2014} and emulsions \cite{Cipelletti2014} show the existence of a critical strain rate and amplitude, at which irreversible particle motion sharply increases, suggesting the presence of a sharp glass-liquid transition in these systems. Nevertheless, first-order transitions in solids imply a change of an underlying symmetry which controls the jump of the order parameter. Such a discontinuous symmetry change, however, has never been observed, and the nature of the transition remains obscure.

Here we resolve this controversy and provide the first experimental evidence that the stress-induced failure of glasses proceeds via a sharp symmetry change in the microscopic degrees of freedom of the glass. Using a combination of x-ray scattering and rheology to directly observe microscopic distortions upon increasing applied strain, we identify an underlying symmetry change in the angular correlation of particles. We observe a sharp symmetry change from anisotropic solid to isotropic liquid structure factor, and relate it to a sudden transition from affine to non-affine displacements. The transition occurs when affine and non-affine displacements reach, respectively, 4 and 0.8$\%$ of the average particle distance; hence, the material reduces strain energies by sharply releasing some of the affine elastic strain into non-affine displacements. We show that the discrete-like anisotropic symmetry is entirely a result of the imposed deformation field and this is likewise reflected in a sharp transition from correlated to uncorrelated (Gaussian) intensity fluctuations, confirming the sharp nature of the transition.
\begin{figure*}[tp]
\centering
{\includegraphics[width=0.32\linewidth]{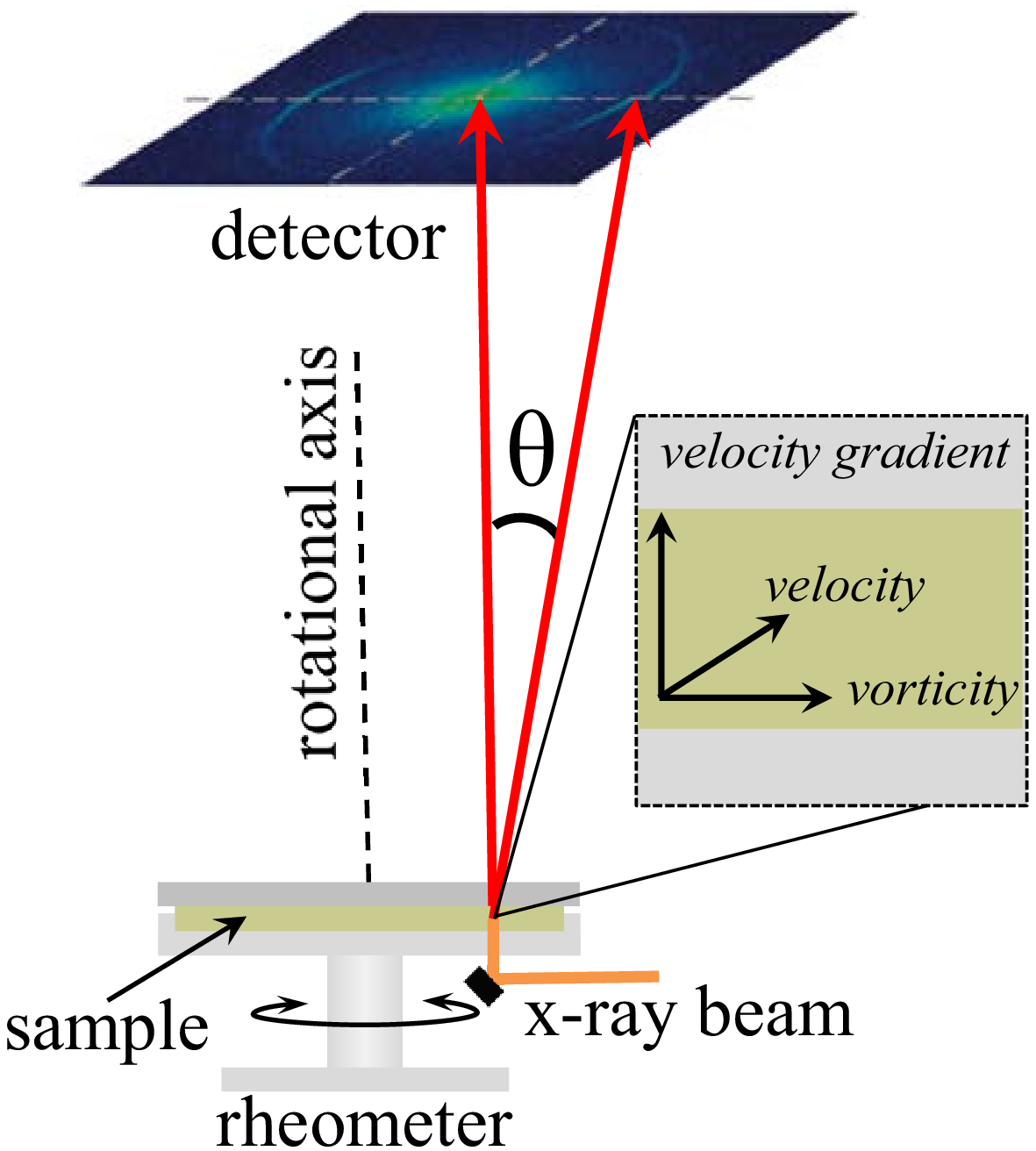}}\quad
\put(-155,0){(a)}
{\includegraphics[width=0.33\linewidth]{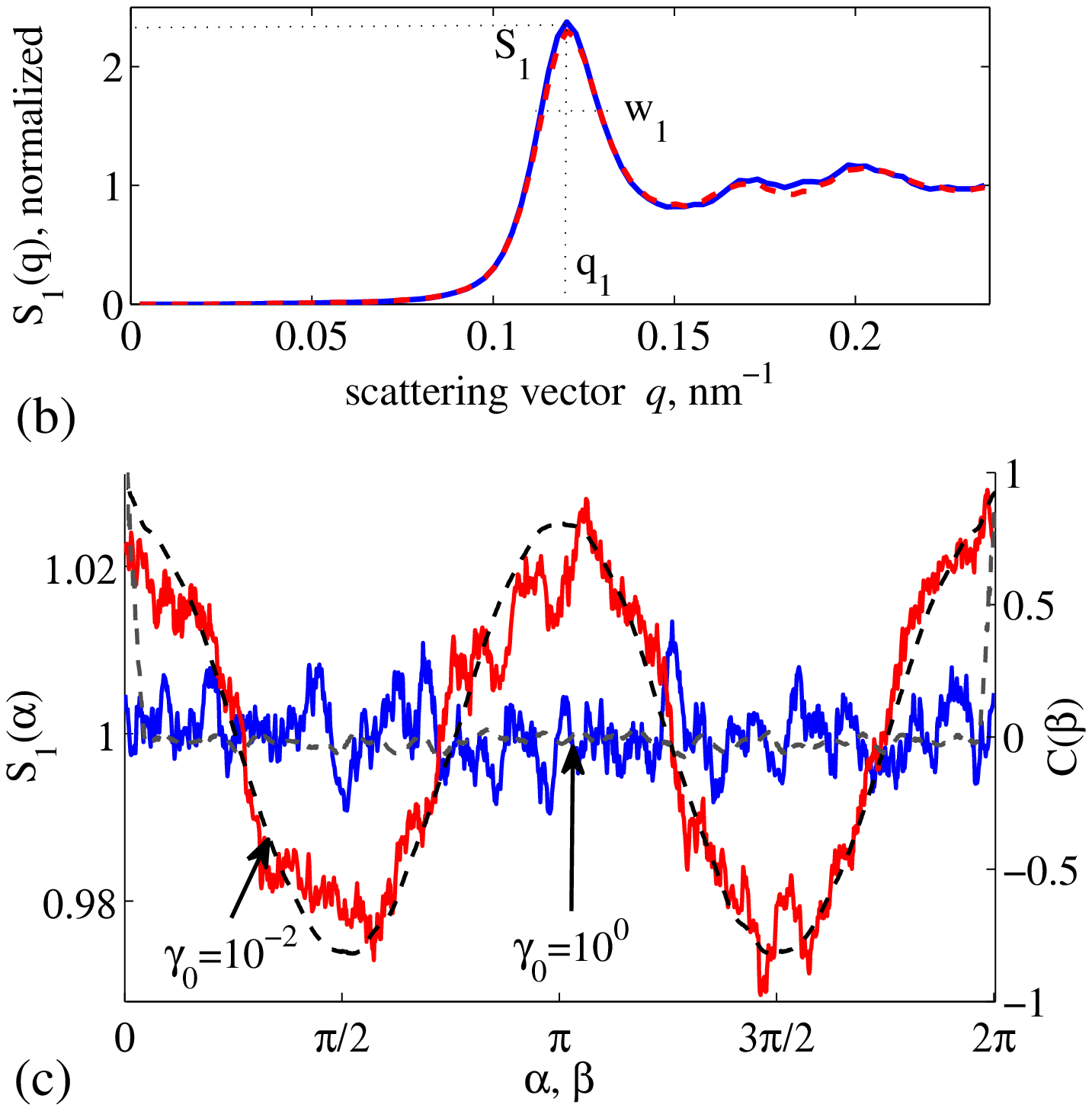}}\quad
{\includegraphics[width=0.26\linewidth]{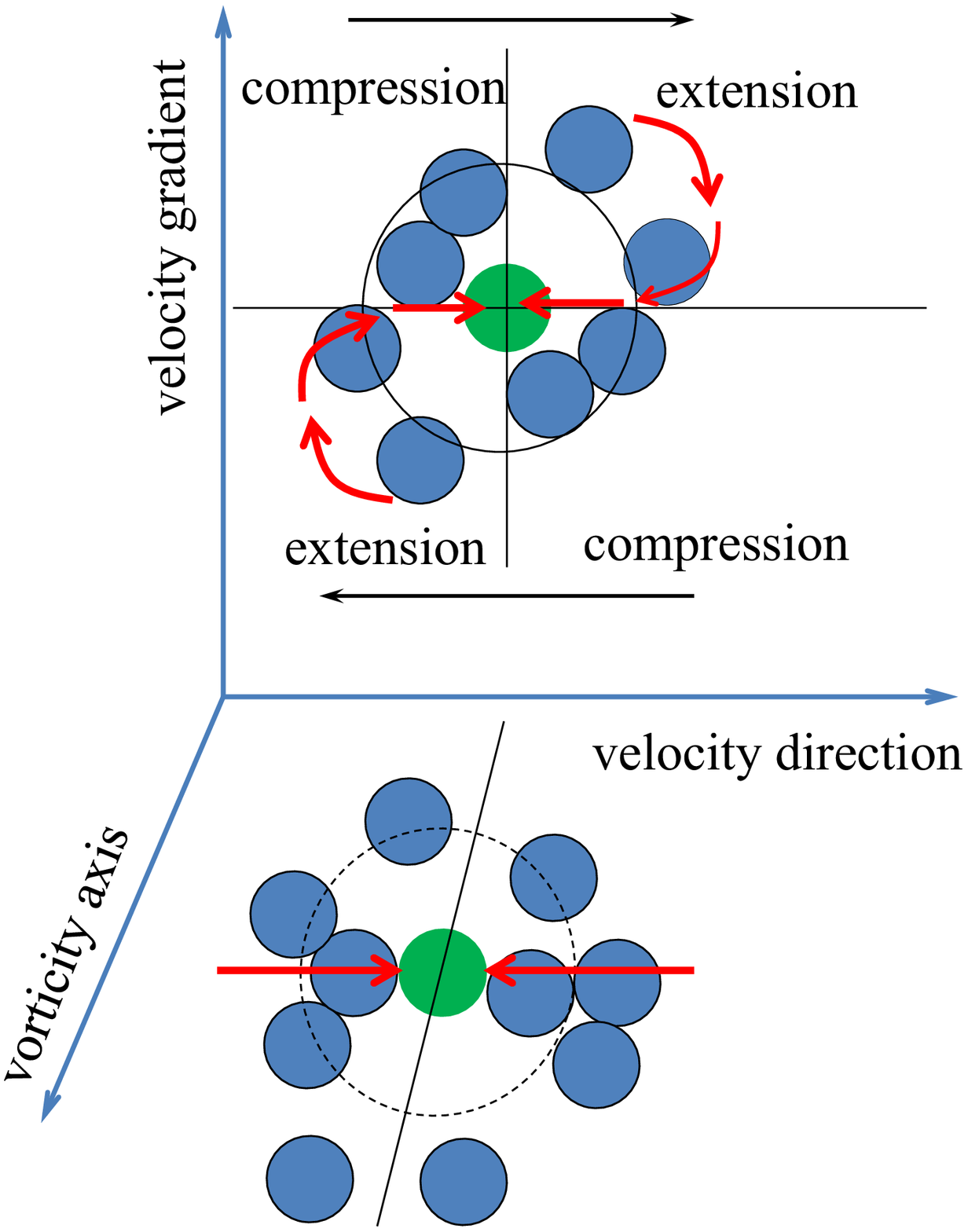}}
\put(-144,0){(d)}
\caption{{\bf X-ray measurement of affine shear distortion of amorphous structure}
(a) Schematic of the experimental setup illustrating the x-ray beam path with respect to the rheometer and the layer of sheared suspension. The rheometer is stress controlled and we use plate-plate geometry. The x-ray beam passes through the suspension at 0.78 times the disc radius; the beam diameter is smaller than 0.1 mm, much smaller than the disc radius of 18mm. The velocity, vorticity and velocity gradient directions at position of the beam is shown in the inset.
(b) Angle-averaged structure factor $S_1(q)$ of the colloidal glass in quiescent state (blue solid curve) and sheared state (red dashed curve).
(c) Angle-dependent height of the first peak of the structure factor shows affine nearest- neighbor distortion for small ($\gamma_0=10^{-2}$, red curve) and no distortion for large strain amplitudes ($\gamma_0=10^{0}$, blue curve). Corresponding angular correlation functions $C(\beta)$ (black dashed lines and axis on the right) quantify the degree of affine distortion. All angles are given relative to the velocity direction.
(d) Schematic of elastic shear distortion of the nearest-neighbor structure. Particles move closer to the central particle in the compression sector, and move further away in the dilation sector of the shear plane. Illustration shows projection in the velocity direction-velocity gradient plane (top) and the velocity-vorticity plane probed by x-ray diffraction (bottom).
\label{fig1}}
\end{figure*}
Our identification of mechanical failure as novel sharp affine-to-nonaffine transition has crucial implications for theories of plastic deformation and applications of amorphous solids in material science and engineering.

\section{Results}
Direct observation of shear-induced distortions in the glass is achieved using  simultaneous oscillatory rheology and synchrotron x-ray scattering. The highly brilliant synchrotron x-ray beam is launched through the sheared layer of suspension from which it scatters~(Fig.~\ref{fig1}a). The particle volume fraction of the suspension is $\phi \sim 0.58$, close to the colloidal glass transition (see Materials and Methods). We apply oscillatory shear at fixed frequency and increasing amplitude; this probes the glass with an increasing oscillatory shear field from the linear elastic to the nonlinear flow regime and allows measurement of the storage and loss moduli, $G'$ and $G''$ simultaneously with the measured structure factor. The angle average of the structure factor, shown in Fig.~\ref{fig1}b, reveals the typical short-range order of a dense glass. When we resolve this structure factor in the diffraction plane, we reveal a characteristic shear-induced distortion of the nearest-neighbor structure as shown in Fig.~\ref{fig1}c:
The twofold (p-wave) symmetry indicates the affine distortion of nearest neighbor configurations under the applied shear deformation. This is illustrated in Fig.~\ref{fig1}d: particles tend to accumulate and crowd in the compression sectors of the shear plane, and dilate in the extension sectors. Because of strong excluded-volume repulsion, particle accumulation and loss do not balance, leading to a net depletion of particles from the cage, and a resulting loss of connectivity of the glass structure. In an elastic material, the particle displacement field is affine, i.e. homogeneous across the material. The affine distortion of nearest neighbor cages induces force-dipoles, directed along the velocity direction, with an associated elastic field that has exactly the twofold symmetry (p-wave symmetry, $\propto\cos^{2}\alpha$) as observed in Fig.~\ref{fig1}c (see Supporting Information). Hence, the two-fold symmetry is consistent with the elastic response of the glass to the applied shear.

\begin{figure*}[tp]
\centering
\includegraphics[width=0.37\linewidth]{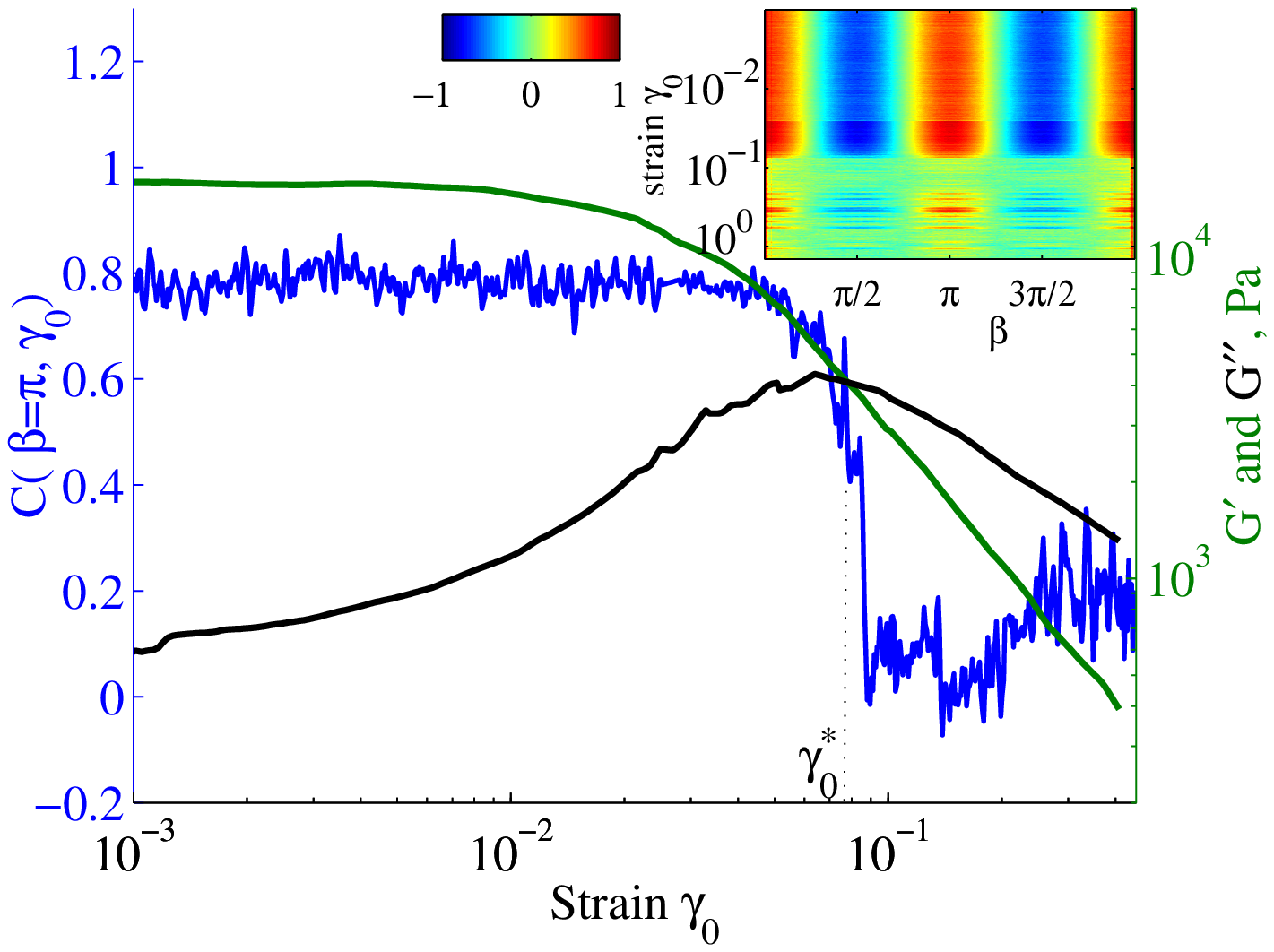} \quad
\put(-190,0){(a)}
\includegraphics[width=0.33\linewidth]{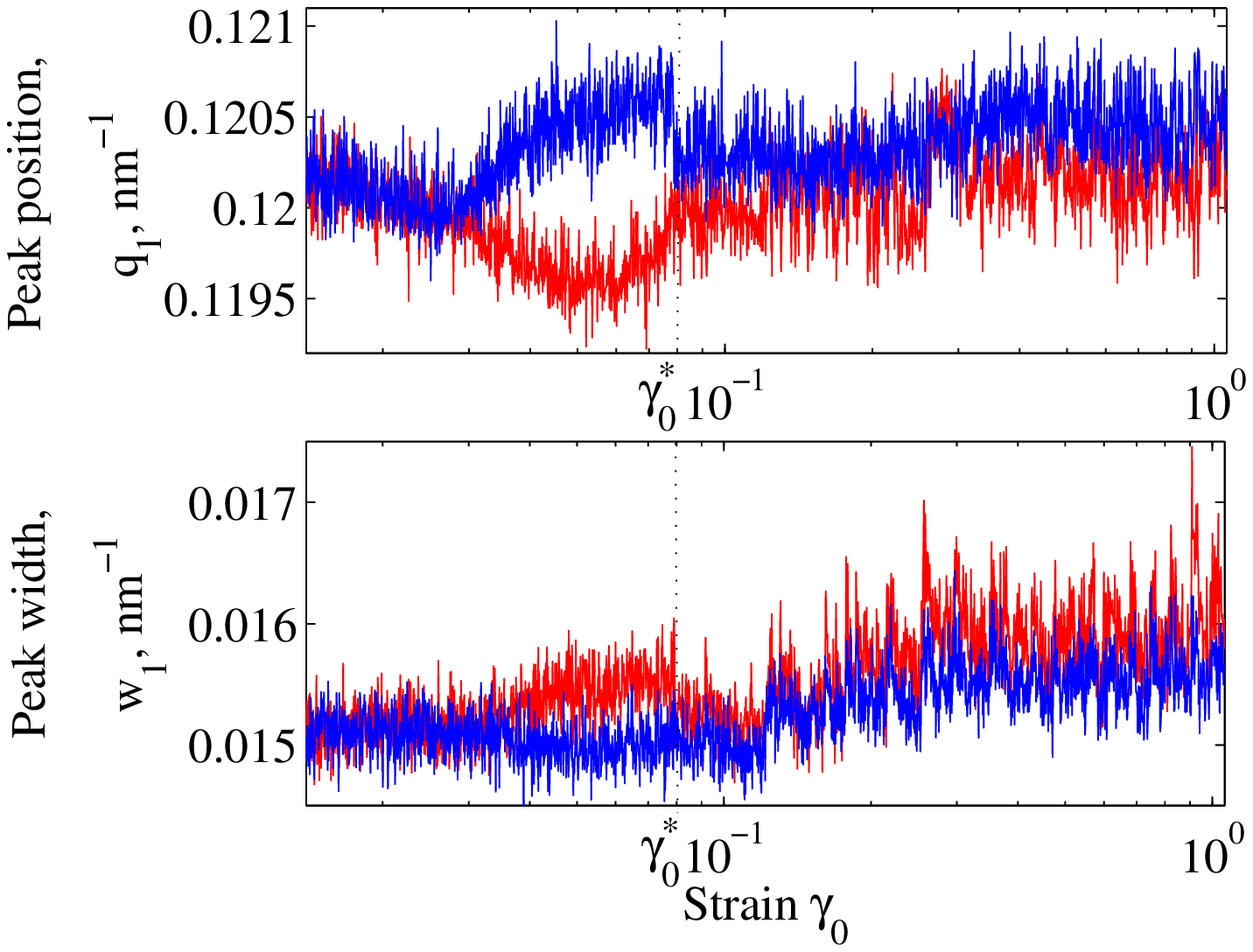} \quad
\put(-165,70){(b)}
\put(-165,0){(c)}
\includegraphics[width=0.24\linewidth]{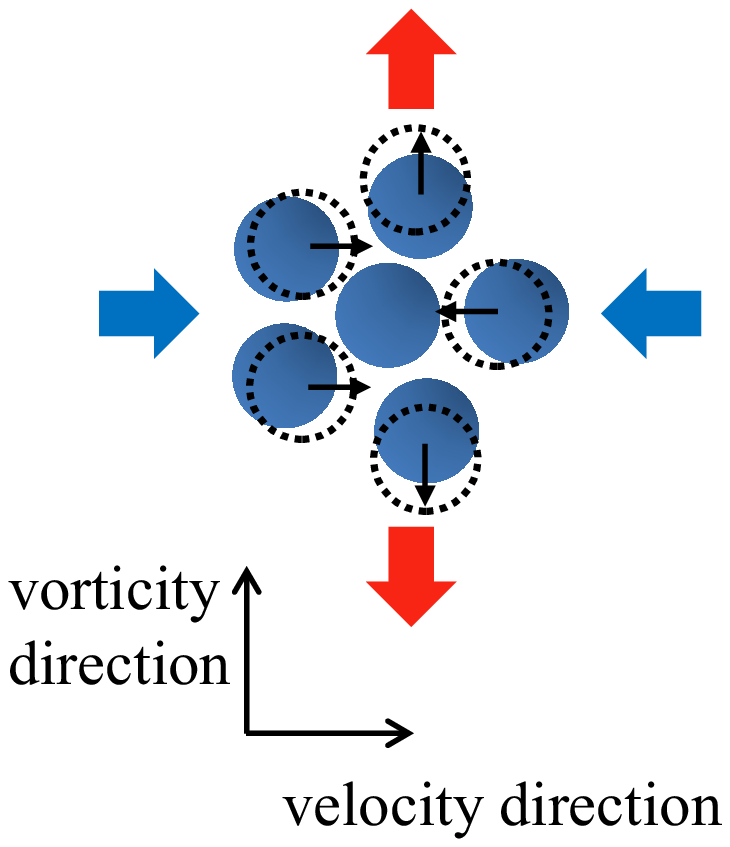}
\put(-110,0){(d)}
\caption{{\bf Sharp transition of the nearest-neighbor structure} (a) Order parameter $C(\beta=\pi,\gamma_0)$ of affine distortions shows sharp drop as a function of strain amplitude (left axis, blue). Also indicated are the elastic and viscous moduli, $G'$ and $G''$ (right axis, green and black). Inset: Contour plot showing the evolution of the angular correlation function $C(\beta,\gamma_0)$ (see color bar) as a function of strain amplitude $\gamma_0$ (vertical axis). (b)-(c) Structural distortion measured from the peak position and width. (b) Position $q_1$ of the first peak of the structure factor indicates compression along (blue) and dilation perpendicular to the velocity direction (red). (c) Peak width $w_1$ as a function of strain. (d) Schematic illustrating the emerging distortion of the nearest-neighbor structure for $\gamma < \gamma_0^{\star}$.
\label{fig2}}
\end{figure*}
This two-fold symmetry vanishes at larger strain: at strain amplitudes of around 10$\%$, the structure factor becomes isotropic (Fig.~\ref{fig1}c), indicating loss of affine distortions. To quantify these affine distortions, we use angular correlations that pick out the characteristic symmetry of the structure factor while averaging over fluctuations, as shown by the dashed lines in Fig.~\ref{fig1}c. These angular correlations hence characterize the degree of elastic response to the imposed shear deformation.
We can now follow the evolution of the structural symmetry as a function of the applied strain. To do so, we define a structural order parameter from the peak value of the correlation function, which is 1 for ideal twofold symmetry, and 0 for complete symmetry loss, and illustrate its evolution in Fig.~\ref{fig2}a. This order parameter drops sharply to zero at $\gamma_0^{\star}\sim 0.08$, indicating the abrupt loss of affine distortions, and the sudden transition to an isotropic liquid-like state. This is demonstrated most clearly in the inset, where we show the full evolution of the angular correlation function along the vertical axis: Red and blue color at small strain (top) indicate the twofold symmetry associated with affine distortions, while green color at larger strain (bottom) indicates the isotropic state. The symmetry vanishes abruptly at $\gamma_0^{\star}$, indicating a sharp loss of orientational order and thus melting in the orientational degrees of freedom. At the same time, the mean value of $S(q)$ does not change, indicating robust translational degrees of freedom. This sharp symmetry change is observed in all other measures of the angle-dependent structure factor such as its peak height and width. It reminds of first-order equilibrium transitions, but in the case here is induced by the applied strain. The observed sharpness is in agreement with a well-known theorem~\cite{Brazovskii1975} that transitions between a (liquid) state with isotropic symmetry (continuous symmetry group) and a (solid) state with broken-symmetry (e.g. the discrete point-group symmetry of a lattice) cannot occur continuously, and have to be first-order in nature. We hence conclude that at $\gamma_0^{\star}\sim 0.08$, the glass exhibits a sharp symmetry change in its microscopic degrees of freedom, induced by the applied oscillatory shear.

This transition results from the increasing distortion of the glass structure associated with the increasing applied strain amplitude. We measure the corresponding non-affine component directly from the difference of nearest-neighbour peak positions in the flow and vorticity direction. In the shear-neutral diffraction plane, affine shear displacements vanish, and any observed distortion must be due to non-affine components. The emerging non-affine distortion, shown in Fig.~\ref{fig2}b, indeed confirms the mechanism proposed in Fig.~\ref{fig1}d: with increasing strain amplitude, the nearest-neighbor peak moves to larger $q$ in the flow direction, and to smaller $q$ in the perpendicular direction. Because of the reciprocal relationship between wave vector $q$ and lengths in real space, this reflects a real-space contraction along the flow direction, and dilation perpendicular to it, as demonstrated schematically in Fig.~\ref{fig2}d. From the difference of orthogonal distances, $\Delta q \sim 0.001~nm^{-1}$, we determine that the relative non-affine distortion is $\Delta q / q \sim 0.8 \%$, while the affine strain is $\sim 4\%$, in good agreement with the value estimated from the correlation function in Fig.~\ref{fig1}c~\cite{footnote1}.
The anisotropy vanishes abruptly at $\gamma_0\sim\gamma_0^{\star}$, again signaling the transition into the liquid-like state: the material can no longer sustain the anisotropic structure, and changes spontaneously into an isotropic fluid-like state. This is likewise reflected in the peak width, whose weak anisotropy vanishes at the transition as shown in Fig.~\ref{fig2}c.

\begin{figure}[tp]
\centering
\includegraphics[width=0.9\linewidth]{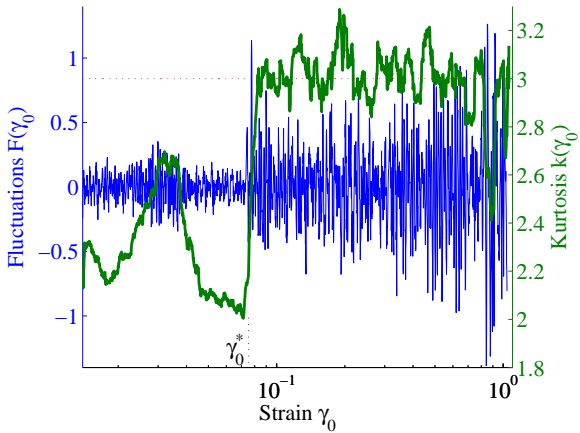}
\caption{{\bf Sharp increase of fluctuations at the glass-liquid transition.}
Normalized fluctuations of the order parameter (left axis, blue curve) and
kurtosis $\kappa$ (right axis, green curve) as a function of the
increasing strain amplitude. At $\gamma_0^*$, fluctuations increase sharply, and the kurtosis jumps to its Gaussian value 3 (red dotted line) indicating liquid-like response. The small peak of $\kappa$ at strains below $\gamma_0^*$ is attributed to combined emergence of anisotropy in peak position and peak width as shown in Fig.~\ref{fig2}b, c. Fluctuations are averaged over one oscillation period (see Supporting Information).
\label{fig3}}
\end{figure}

Further signature of the transition is obtained in the fluctuations of the scattered intensity. We immediately see that at $\gamma_0^{\star}$, the amplitude of fluctuations increases abruptly as demonstrated in Fig.~\ref{fig3} (blue). To further investigate the nature of fluctuations, we distinguish correlated and uncorrelated fluctuations using the kurtosis $\kappa$: a kurtosis value of 3 indicates a Gaussian distribution and thus uncorrelated fluctuations, while a kurtosis value smaller than 3 indicates correlated fluctuations. We compute $\kappa$ for each strain from the intensity distribution along the first diffraction ring (see Materials and Methods) and show its evolution in Fig.~\ref{fig3}. Concomitantly with the increase in amplitude, the nature of fluctuations changes: A sharp rise of $\kappa$ to a value of $\sim 3$ indicates a sudden transition from correlated to uncorrelated Gaussian fluctuations. Such Gaussian fluctuations are indeed indicative of diffusive particle transport in liquids, giving yet another evidence of the melting of the glass in the microscopic degrees of freedom. Together, the disappearance of anisotropy, the increase of fluctuation amplitude, and the transition to Gaussian distributions conclusively demonstrate the sudden melting, i.e. sharp transition from a solid to a liquid-like state of the amorphous material under the applied shear.

To relate this transition to the macroscopic mechanical properties of the glass, we use the stress response recorded simultaneously to determine the mechanical moduli. In linear response, the rheology is characterized by the storage and loss moduli, $G^\prime$ and $G^{\prime\prime}$, determined from the in and out-of-phase harmonic stress response. We show these moduli as a function of strain together with the structure factor in Fig.~\ref{fig2}a. They exhibit the well-known strain dependence of soft glassy materials: the dominance of $G^\prime$ over $G^{\prime\prime}$ at small strain indicates the predominant elastic response of the material, and their decrease and intersection at higher strain indicate the loss of elasticity and flow of the glass. The intersection of $G^{\prime}$ and $G^{\prime\prime}$ has been associated with the transition from a solid to a liquid state of the material~\cite{BonnDenn}. Remarkably, this intersection coincides precisely with the sharp structural transition, allowing us to identify it with the rheological solid-to-liquid transition of the material~\cite{footnote2}. We show in the supporting information that the macroscopic criterion $G^{\prime} = G^{\prime\prime}$ is in fact equivalent to the equality of the two central microscopic contributions: affine and non-affine~\cite{Zaccone11}. We hence find experimentally that the rheological yielding of the material occurs precisely when the affine and non-affine manifestations of the shear modulus become equal. These results indicate that the affine to non-affine transition is {\it the} central microscopic mechanism that controls the solid-liquid transition of the amorphous material.
We note that the moduli $G^\prime$ and $G^{\prime\prime}$ represent only the first harmonic response; higher harmonics are not taken into account, although they can be quite significant~\cite{Wilhelm,Rogers2012}. The inclusion of these higher harmonics might evidence a sharper transition also in the mechanical response. Indeed, recent theoretical work in systems with excluded-volume interactions~\cite{Zaccone14} indicates that the shear modulus exhibits a sharp transition (as also suggested by Mode-Coupling Theory~\cite{Brader2010}), but the jump is very small and may not be easily resolved in experiments, thus leaving the overall impression of a continuous transition. Such continuous transition is recently also observed in slow, continuous shear~\cite{Friedman2012,Antonaglia2014}. Our result of a sharp transition in oscillatory experiments that rather probe the stress stability limit of the material~\cite{YieldingCollGlass} is particularly interesting for a full universal understanding of the mechanical behavior of glasses.

In conclusion, the abrupt change of structural symmetry and nature of fluctuations consistently demonstrate a sharp transition in the microscopic degrees of freedom of a glass upon mechanical failure. The shear-induced, discrete symmetry characteristic of elastic solids vanishes abruptly, and isotropic Gaussian fluctuations characteristic of liquids appear, indicating a surprisingly sharp, dynamically-induced transition (symmetry-breaking) from a solid to a liquid-like state. This sharp transition and symmetry-breaking are all hallmarks of first-order equilibrium transitions. Furthermore, the symmetry breaking is an essential ingredient in the context of generalized rigidity and Goldstone's theorem~\cite{AndersonBook}, to explain the emergence of stiffness from the breaking of the isotropic symmetry of the liquid. This symmetry change has not been detected before which made it impossible, so far, to explain why glasses are, at all, rigid.
We reveal this symmetry-breaking here as dynamically induced (as opposed to spontaneous),
induced by the applied oscillatory shear field. While this point requires further theoretical investigation, we identify this transition as fundamental affine to non-affine transition in the microscopic degrees of freedom: The increasing shear amplitude causes increasing distortion of the nearest-neighbor structure, leading to loss of connectivity and proliferating nonaffine displacements, until at crossover to flow, the response becomes completely dominated by non-affine displacements.
We suggest that, in analogy to first-order equilibrium transitions, this sharp transition can result from two underlying free energy curves due to the two competing modes of displacements - affine and non-affine. The material then exhibits a sharp jump in the contributions of the two respective modes of displacement, consistent with the order parameter jump from initially $\sim 0.8$, suggesting the presence of both contributions. While our combined x-ray scattering and rheology measurement allows us to clearly observe this transition in a colloidal glass, we expect it to be a ubiquitous feature of the mechanical failure of amorphous solids, and to occur in a wide range of technologically important materials from soft colloidal and polymer to hard molecular glasses, where the hard-core repulsion is replaced by the quantum mechanical short-range repulsion due to Pauli's principle. Our results hence suggest a new unified framework of the mechanical stability and failure of amorphous materials crucial to their application as engineering materials, and offers new perspectives on the glass transition.

\section{Methods}

\subsection{Experimental setup and colloidal samples}
The experiments were carried out at the beamline P10 of the synchrotron PETRA III at DESY. To measure the rheology and structure factor simultaneously, we placed an adapted commercial rheometer (Mars II, Thermo Fisher) into the x-ray beam path of the synchrotron. The well-collimated x-ray beam (wavelength $\lambda = 0.154$ nm) is deflected vertically to pass the layer of suspension in the shear-gradient direction, see Fig.~\ref{fig1}a. The suspension consists of silica particles in saline water (1 mM  NaCl to screen the particle charges), with a diameter of $50$ nm and a polydispersity of $10\%$ preventing crystallization.
Dense samples with effective volume fraction $\phi \sim 0.58$ at the colloidal glass transition were prepared by diluting centrifuged samples. Measurements of the relaxation time yielded $\tau \sim 10^6 t_B$~\cite{Denisov2013}, with $t_B$ the relaxation time at infinite dilution, consistent with $\phi \sim 0.58$~\cite{vanMegen1998}.
After loading, the samples are sealed with low-viscosity oil to prevent evaporation and guarantee sample stability over more than 4 hours, allowing us to measure samples repeatedly and reproducibly. Samples were initialized by a fixed protocol (preshear at $\dot{\gamma}=0.1$ s$^{-1}$ for 120 seconds, followed by 600 seconds rest). We apply oscillatory strain with frequency $f=1$ Hz and amplitude $\gamma_0$ increasing from $\gamma_{0min} = 10^{-4}$ to $\gamma_{0max} = 1$ (100 points on a logarithmic scale, three oscillations averaged per cycle, leading to total duration of the experiment of around 5 minutes).

\subsection{X-ray data acquisition and analysis}
We use a Pilatus detector at a distance of $D = 280$ cm and a frame rate of 10 Hz to measure the scattered intensity in the velocity-vorticity plane. The
detector (pixel size $172 \times 172$ $\mu$m$^2$) covers scattering angles $\theta$ between $0.03$ and $0.5^{\circ}$, allowing access to wave vectors $q = (4\pi/\lambda) \sin(\theta/2)$ in the range $qr_0 = 0.5$ to $5$. From the recorded intensity, we determine the structure factor $S(\textbf{q})$ by subtracting the solvent background and dividing by the particle form factor determined from dilute suspensions. We focus on the first peak of the structure factor to investigate the nearest-neighbor structure.

\subsection{Angular correlation analysis}
We use angular correlation functions to investigate the symmetry change of the structure factor. Angular correlations are computed on the first diffraction ring $S_1(\alpha)$ according to
\begin{equation}
C(\beta)=
\frac{\int_0^{2\pi}(S_1(\alpha+\beta)-<S_1(\alpha)>)(S_1(\alpha)-<S_1(\alpha)>)d\alpha}
{\int_0^{2\pi}(S_1(\alpha)-<S_1(\alpha)>)^2 d\alpha}.
\label{eq1}
\end{equation}
Here,  $\alpha$ and $\beta$ are polar angles in the diffraction plane, and we integrate over the angle $\alpha$ as a function of the correlation angle $\beta$. Possible effects of elliptical distortion of the first ring are reduced by averaging radially over a range of wave vectors ($\Delta q \sim2w_1$) around $q_1$. We define the peak value $C(\beta = \pi)$ as structural order parameter; this allows us to measure the symmetry change as a function of applied strain.

\subsection{Fluctuation analysis}
To analyze the fluctuations of the order parameter and structure factor, we first check whether there is a characteristic time scale of fluctuations. Such time scale can for example be the underlying oscillation period, during which the glass may yield and reform~\cite{Rogers2011b}. To check for such time scale, we compute time correlations according to
\begin{equation}
F(\Delta t)=
\frac{1}{T}\int_0^{T}(C(t+\Delta t)-<C(t)>)(C(t)-<C(t)>)dt,
\label{eq2}
\end{equation}
where $t\sim \log(\gamma_0/\gamma_{0min})$ and we correlate order parameter values $C(t)=C(\beta=\pi,t)$ as a function of delay time $\Delta t \sim \Delta\log(\gamma_0)$. For sufficiently large averaging time interval, $T$, the time correlation should pick out the typical fluctuation time scales if present. However, our data gives no evidence of a characteristic time scale (see Supporting Information), as we have independently verified by Fourier analysis. We thus interpret the fluctuations as noise. To obtain the data shown in Fig.~\ref{fig3} (blue), we average over all ten points within the oscillation cycle, i.e. we choose $T$ to be the oscillation period.

To investigate the nature of fluctuations, we finally compute the kurtosis $\kappa = m_4/m_2^2$, where $m_4$ and $m_2$ are the fourth and second moment of the intensity distribution. To follow $\kappa$ as a function of strain, we compute instantaneous values of $\kappa$ from the intensity distribution along the first ring. We define the i-th moment $m_i=\frac{1}{2\pi}\int_{0}^{2\pi}{(S_1(\alpha)-<S_1(\alpha)>)^i d\alpha}$, where we integrate over the polar angle $\alpha$ in the diffraction plane. This allows us to follow $\kappa$ instantaneously as a function of strain.

\section{Acknowledgements}
We thank DESY, Petra III, for access to the x-ray beam. This work was supported by the Foundation for Fundamental Research on Matter (FOM) which is subsidized by the Netherlands Organisation for Scientific Research (NWO). P.S. acknowledges support by Vidi and Vici grants from NWO.

\end{document}